\documentstyle[preprint,aps]{revtex}

\begin{document}
\bibliographystyle{prsty}
\draft

\title{\large\bf Experimental Test of Complementarity by Nuclear Magnetic Resonance
Techniques}
\author{Xiwen Zhu$^{1}$
\thanks{Author to whom correspondence should be addressed. Email address: xwzhu@nmr.whcnc.ac.cn}, 
Ximing Fang$^{2,1}$, Xinhua Peng$^{1}$, Mang Feng$^{1}$,
Kelin Gao$^{1}$, and Fei Du$^{1}$}
\address{$^{1}$ Laboratory of Magnetic Resonance and Atomic and Molecular Physics, \\
 Wuhan Institute of Physics and Mathematics, Chinese Academy of Sciences,\\
Wuhan 430071, People's Republic of China\\
$^{2}$ Department of Physics, Hunan Normal University, Changsha, 410081, People's Republic of China}

\maketitle

\begin{abstract}

 We have tested complementarity for the ensemble-averaged spin
states of nuclei $^{13}$C in the molecule of $^{13}$CHCl$_{3}$ by the use of the spin
states of another nuclei $^{1}$H as the path marker. It turns out that the
wave-particle duality holds when one merely measures the probability density
of quantum states, and that the wave- and particle-like behavior is
simultaneously observed with the help of measuring populations and
coherences in a single nuclear magnetic resonance(NMR) experiment. Effects
of path-marking schemes and causes of the appearance and disappearance of
the wave behavior are analyzed.
\end{abstract}
\vskip 1cm

\narrowtext

Whether the wave-like and particle-like behavior, the fundamental attributes
of quantum mechanical entities, can display or be observed simultaneously in
a single experiment has been argued[1] since the foundation of quantum
mechanics in 1920's. Bohr claimed the answer to be negative and expressed it
as complementarity[2], one of the most basic principles of quantum
mechanics. Bohr complementarity is usually illustrated by the famous
two-slit experiment, in which a quantum object reaches a final state along
two different paths simultaneously. The probability density or population of
the final state displays interference when two paths are indistinguishable.
Any attempt or practice to gain 'which-path' information unavoidably
destroys the interference pattern. As the different path and the
interference pattern are commonly attributed to a particle and a wave
correspondingly, the results stated above are thus superficially referred to
unobservableness of the particle- and wave-like behavior under the same
experimental condition, in short, wave-particle duality of matter.

Complementarity was justified by analyzing the physical principles of early
'thought' experiments like Einstein's recoiling slit[1] or Feynman's
light microscope[3] and recent experimental results with two trapped ions[4], 
an atom interferometer[5] and a pair of entangled photons[6] inspired
by the proposal of Scully $et~al$[7]. The mechanisms or origins that enforce
complementarity, however, vary from one experimental situation to another[3,5,7-11]. 
In the 'thought' experiments with classical which-path
markers, the act of labelling or measuring the paths invariably introduces a
random momentum transfer or uncontrolled phase shift, thus washing out the
interference fringes, which can be explained by Heisenberg uncertainty
relation. While marking the path quantum-mechanically in the recent
experiments does not lead to noticeable phase shift, disappearance of
interference when measuring the probability density of the final state
originates in quantum correlations of the which-path detector and observed
quantum state, and the uncertainty principle plays no role.

In all previous variants of the two-slit experiment the wave behavior was
observed by measuring the population in the final state, and the quantum
state of the center-of-mass motion for micro-particles was investigated. Is
complementarity applicable to the internal states for a microscopic object
or even a bulk ensemble? What are the effects of different path-marking
schemes and the causes of appearance and disappearance of the wave behavior?
Is there another way to detect the wave behavior of the quantum states
except the population measurement? Most significantly, is observation of the
wave and particle behavior under the same experimental condition absolutely
prohibited indeed? These questions motivated us to perform this study by
using an ensemble of nuclear spins as sample with NMR techniques.

Our experiment scheme is illustrated in Fig.1. An ensemble of molecules with
two spin-$\frac {1}{2}$ nuclei is firstly prepared in the initial state $\left| \psi
_i\right\rangle =\left| 0\right\rangle _b\left| 0\right\rangle _a$ $\equiv
|00\rangle $, where $b$ is the observed nuclei, $a$ is the labelling ones,
and $\left| 0\right\rangle $ and $\left| 1\right\rangle $ represent the
states with spin-up and spin-down respectively. In Fig.1a, an operation 
$R_i^b(\theta )= \pmatrix{\alpha (\theta )& -\beta (\theta ) \cr 
\beta (\theta ) & \alpha (\theta )}$
on nuclei $b$ transforms the initial state 
$\left| 00\right\rangle$ into an intermediate state 
$|\psi _{1m}> =\left[ \alpha
(\theta )\left| 0\right\rangle _b+\beta (\theta )\left| 1\right\rangle
_b\right] $ $\left| 0\right\rangle _a$(assuming $|\alpha (\theta )|^2+|\beta
(\theta )|^2=1$ ), which is then transferred to the final state 
$|\Psi_{1f}>=\frac {1}{\sqrt{2}}[(\alpha(\theta)+\beta(\theta)e^{i\phi})|0>_{b}
+(\beta(\theta)-\alpha(\theta)e^{-i\phi})|1>_{b}]|0>_{a}$ by the
operation $U^b(\phi )=\frac 1{\sqrt{2}}\pmatrix{1 & e^{i\phi } \cr 
-e^{-i\phi } & 1}$
on $ b$. As the labelling nucleus  $a$ remains in the state 
$|0> _a$ all the time as a spectator, two paths along which the nuclei 
 $b$ reaches the final state $|0\rangle _b(|1\rangle _b)$ through the
intermediate state $|0\rangle _b$ and $|1\rangle _b$ are indistinguishable.
The probability density of finding {\it b} in the final state $|0\rangle
_b(|1\rangle _b)$ is measured to be $\frac 12\left[ 1+2\alpha (\theta )\beta
(\theta )\cos \phi \right] (\frac 12\left[ 1-2\alpha (\theta )\beta (\theta
)\cos \phi \right] )$. Repeating experiment by changing $\phi $ will produce
Ramsey fringes denoted by the term $\pm \alpha (\theta )\beta (\theta )\cos
\phi $ , showing wave behavior obviously[12]. While in Fig.1b, the
operations $R_2^{ba}(\theta )=CN_{ba}\bullet R_1^b(\theta )$ and $U^b(\phi ) 
$ are successively executed, with $CN_{ba}$ being the controlled-NOT gate in
quantum computing, $b$ and $a$ , the control and target qubits respectively.
The intermediate states $|\psi _{2m}\rangle =\alpha (\theta )|00\rangle
+\beta (\theta )|11\rangle $ and the final state $|\psi _{2f}\rangle =\frac 1%
{\sqrt{2}}\{|0\rangle _b\left[ \alpha (\theta )|0\rangle _a+\beta (\theta
)e^{i\phi }|1\rangle _a\right] +|1\rangle _b\left[ -\alpha (\theta
)e^{-i\phi }|0\rangle _a+\beta (\theta )|1\rangle _a\right] \}$ are thus
obtained. Since the states $|0\rangle _b$ and $|1\rangle _b$ are definitely
marked by $|0\rangle _a$ and $|1\rangle _a$ respectively in the preparation
of the intermediate state, i.e., $|0\rangle _b\longleftrightarrow |0\rangle
_a$ and $|1\rangle _b\longleftrightarrow |1\rangle _a$ , and those markers
remain unchanged in the later transformation $U^b(\phi )$, two paths along
which spin $b$ involves to the final state $|0\rangle _b(|1\rangle _b)$
from the intermediate states $|0\rangle _b$ and $|1\rangle _b$
simultaneously can be precisely identified by the states of $a$. The
probability densities of finding spin $b$ in the final state $|0\rangle
_b$ from the intermediate states $|0\rangle _b$ and $|1\rangle _b$ are $%
\frac 12\left| \alpha (\theta )\right| ^2$ and $\frac 12\left| \beta (\theta
)\right| ^2$ respectively ( the same for the final state $|1\rangle _b$ ).
The particle behavior is thus clearly revealed. However, the probability
density of finding $b$ in the final state $|0\rangle _b$ or $|1\rangle
_b $ is to be 1/2, a constant independent on $\phi$ , due to the
orthogonality of the quantum correlation states, so Ramsey fringes
disappear. If one can detect the coherence, $\alpha (\theta )\beta (\theta
)\sin \phi $ , between the final states $|00\rangle $and $|01\rangle $(
similar for $|10\rangle$ and $|11\rangle$) while measuring the probability
density of those states, then the wave behavior reappears via the dependence
of the coherence on $\phi $ , and the particle behavior is shown by the path
marker and measurement of the population from different paths. Therefore,
simultaneous observation of the wave and particle behavior in a single
experiment is accomplished.

The sample we used is the molecule of carbon-13 labelled chloroform 
$^{13}$CHCl$_3$, nuclei $^{13}$C and $^1$H of which were chosen as observed
 nuclei $b$ and labelling ones $a$ respectively. Experiments were carried out
on a Bruker ARX 500 spectrometer with conventional liquid NMR technique[13]. 
The resonance frequencies are about 125 MHz for $^{13}$C and 500 MHz
for $^1$H, and the scalar coupling constant $J_{ab}$ is near 215 Hz. Four lines,
two for $^{13}$C and two for $^1$H spectra, can be well resolved. By
applying two simultaneous line-selective pulses with appropriate frequencies
and rotation angles as well as a magnetic field gradient pulse consecutively on the
ensemble of nuclei $^{13}$C and $^1$H in thermal equilibrium[14], we
prepared the quantum ensemble in an effective pure state expressed by the
density matrix $\rho _{00}^i=AE+B$ $|00\rangle \langle 00|$, where E is a 
$4\times 4$ unit matrix, A and B are constants with $A\gg B$. As 
$\rho _{00}^{i}$
has the same quantum-mechanical properties and NMR experimental results as
the pure state $|00\rangle$[15,16], the conclusions extracted from
experiments on $\rho _{00}^{i}$ can be applied to characterize $|00\rangle $.
Operation $R_1^b(\theta )$ was implemented by a pulse $(\theta )_{-y}^b$, a
pulse applied on $^{13}$C and rotating a angle $\theta $ along the negative
y-axis. The amplitudes of the wave function of the intermediate state thus
obtained are $\alpha (\theta )=\cos (\theta /2)$ and $\beta (\theta )=\sin
(\theta /2)$. Two values of $\theta $ were set at 90$^{\circ }$ and 53.24$%
^{\circ }$ in different runs, corresponding to the ratio $\left| \alpha (\theta
)\right| ^2/\left| \beta (\theta )\right| ^2$ of populations in two
intermediate states to be 1 and 4. The controlled-NOT gate $CN_{ba}$ was
achieved by a pulse sequence $(\pi /2)_{-y}^a(1/2J_{ab})(\pi /2)_y^{a,b}(\pi
/2)_{-x}^{a,b}(\pi /2)_{-y}^b$ , where $1/2J_{ab}$ is the time interval of
free evolution period[17]. Transformation $U^b(\phi )$ could be denoted by $%
e^{-iI_x^b\theta _1}e^{-iI_y^b\theta _2}e^{-iI_x^b\theta _1}$ with $%
I_{x,y}^b $ being the spin operators of the nuclei $^{13}$C and $\theta
_1=\tan ^{-1}(-\sin \phi )$ , $\theta _2=2\sin ^{-1}(-\cos \phi /\sqrt{2})$
, so it was performed by a pulse sequence $(\theta _1)_x^b(\theta
_2)_y^b(\theta _1)_x^b$ . Various values of $\phi $ were obtained by
assuming appropriate $\theta _1$ and $\theta _2$. So the wave and particle
behavior in given situation could be examined by inspecting the dependence
of populations and coherences on $\phi $ . Reading-out pulses $(\pi /2)_y^a$
and $(\pi /2)_y^b$ were finally applied and spectra for $^1$H and $^{13}$C
were recorded. As the population difference and coherence of relevant states
were transferred to the measured transverse magnetization after applying
read-out pulses and embodied in the real and imaginary parts of the line
intensity of the recorded spectra respectively[13], the normalized
population and coherence of the interested states were extracted in Figs.2-4
by data analyzing and fitting.

Variations of the normalized populations $\rho _{00}^{f_1}$ and $\rho
_{10}^{f_1}$ of the states $|00\rangle $ and $|10\rangle $ versus $\phi $
when two paths along which the spin ensemble of nuclei $^{13}$C involve from
the intermediate states to the final states cannot be discerned by the
invariable state $|0\rangle _a$ of the spectator nuclei $^1$H are shown in
Fig.2. The range of $\phi$ is assumed in $(0,2\pi )$ due to its $2\pi $
periodicity. Ramsey fringes can be clearly seen for two sets of data in the
case of the ratio of the normalized populations in two intermediate states
to be 1 and 4. The visibility of fringes decreases with the population ratio
increased. These experimental results are in good agreement with the
theoretical expectation. Experimental errors are mainly due to the
imhomogeneity of the RF field and static magnetic field of the spectrometer,
imperfect calibration of applied pulses (esp., the $\theta _1$ and $\theta
_2 $ setting in different runs), and signal decay during the experiment[17].

Variations of normalized populations versus $\phi $ when two paths from the
intermediate to final states are distinguishable by the states of the
labelling nuclei $^1$H are depicted in Fig.3. In this case the populations, $%
\rho _0^{f_2}$ and $\rho _1^{f_2}$, of the observed nuclei $^{13}$C in the
final states $|0\rangle _b$ and $|1\rangle _b$ respectively don't vary with $%
\phi $ regularly, i.e., the wave behavior does not manifest itself in this
way. Experimental data are also fairly consistent with the theory.

It can be concluded from Figs.2 and 3 that the conventional description of
Bohr complementarity still holds for the quantum entity characterized by the
internal states of a spin ensemble. The wave behavior displays in the
measured populations when both paths are indistinguishable, and disappears
in population measurements if one attempts to get which-path information.
Nevertheless, we have measured the coherences, $C_0$ and $C_1$, between the
relevant state pairs $(|00\rangle ,|01\rangle )$ and $(|10\rangle
,|11\rangle )$ respectively, besides the population measurements when both
paths were labelled by the states of nuclei $^1$H. The measured coherences of
the final states $|0\rangle _b$ and $|1\rangle _b$ (see Fig.4) also
showed oscillatory property with $\phi $, thus revealing the wave behavior
in another way. Because the results given in Figs.3 and 4 were observed in
the same time and under the same experimental situation, we then conclude
that simultaneous observation of the wave and particle behavior of a quantum
entity has been achieved in a single experiment.

Some remarks will be made as follows. In all previous two-slit experiments
the wave behavior was solely observed through the population measurement.
The results extracted from those experiments, including ours shown in Figs.2
and 3, confirmed that the oscillatory feature of populations excludes with
the path distinguishability. Complementarity stated as "simultaneous
observation of the wave and particle behavior is prohibited" is therefore
valid in and only in the sense that one merely observes the population in
his experiment.

As well known, observation of the wave behavior in classical and quantum
physics requires that the relative phase between relevant wave components
keep constant. The wave behavior can be manifested through the existence of
coherence and population oscillation. As labelling different evolution paths
by means of appropriate quantum states does not bring about any random phase
shift and makes the phase difference between the wave function of the states
involved unchanged, the wave character maintains after path-labelling.
Although the populations in the final states remain invariable with $\phi $,
the oscillating coherence between states originated from different
intermediate states and marked by labelling nuclei $^1$H can surely be used
to show the wave character. Therefore, realization of simultaneous
observation of the wave and particle behavior of any quantum object with the
help of quantum path marking and measuring populations and coherence is
physically reasonable and universal in principle. This scheme is readily
feasible for the internal state of quantum systems using the techniques
existing nowadays or available in the near future, e.g., for the internal
state of atoms and molecules by optical spectroscopy. If some means of
observing the coherence or other observable characterizing the wave behavior
of the center-of-mass states of any quantum entity could be found out and
executed in practice, simultaneous observation of the wave and particle
behavior is then applicable to all quantum systems.

Besides the wave-particle duality, complementarity sometimes has endowed
with more general concept. Two observable are referred to be complementarity
if precise knowledge of one of them implies that all possible outcomes of
measuring the other one are equally possible[7]. All canonical conjugate
pairs (e.g., position and momentum) and two incommutable observable (e.g.,
longitudinal and transverse polarizations) are such examples.
Complementarity in this general sense is consistent with current framework
of quantum mechanics but beyond the scope of the present paper. 

In conclusion, we have experimentally tested Bohr complementarity for the
quantum entity characterized by the internal states of a nuclear spin
ensemble in the bulk matter using concepts of the controlled-NOT gate and
entangled states in quantum computing as well as techniques in NMR spectroscopy.
Simultaneous observation of the wave and particle-like behavior of the
quantum object was realized for the first time. The reasons why both
behaviors could not be observed in the same time were revealed. The
condition and method of observing the both in a single experiment were
proposed. It was indicated that the method proposed in the present paper
could be applied to the internal and center-of-mass states of a single
micro-particle and macroscopic ensemble and would be nearly feasible for the
internal state of some other quantum objects.

The authors thank  Xi'an Mao, Maili Liu, Shibiao Zheng, Tao Zhang and Hansheng 
Yuan for discussions and technical assistance. The work is supported by 
National Natural Science Foundation of China.

\begin{center}{\bf Captions of the figures}\end {center} 

Fig.1~ Schematic of a 'two-slit' experiment with internal states of a
quantum entity. The entity involves from the initial state $|00\rangle$ via
the intermediate to final one by transformations of (a) $R_{1}^{b}(\theta )$
and $U^{b}(\phi )$ ; (b) $R_{2}^{ab}(\theta )$ and $U^{b}(\phi )$ (see
text). 

Fig.2~ Normalized population vs. the relative phase $\phi$ when
evolution paths are indistinguishable. Experimental data points $\bigcirc $
and $\bigtriangledown $ denote the populations $\rho _{00}^{f_{1}}$ in the
state $|00\rangle $, $*$ and $+$ denote $\rho _{10}^{f_{1}}$ in the state 
$|10\rangle$. Theoretical curves expressed by $\frac {1}{2}(1\pm\sin\theta\sin\phi)$ 
are depicted with the solid
lines. (a) $\theta =90^{\circ }$ and (b) $\theta =53.24^{\circ }$ .

Fig.3~ Normalized populations vs. the relative phase $\phi$ when two paths are
distinguishable. Data points $\bigcirc$, $*$, $\bigtriangledown$ and $+$ 
denote $\rho_{00}^{f_2}$, $\rho_{01}^{f_2}$, $\rho_{10}^{f_2}$ and $\rho_{11}^{f_2}$, 
the notations of square and star 
denote $\rho_{0}^{f_2}$ and $\rho_{1}^{f_2}$ respectively. Theoretical curves expressed 
by $\rho_{00}^{f_2}=\rho_{10}^{f_2}=\frac {1}{2}\cos^{2}\frac {\theta}{2}$, 
$\rho_{01}^{f_2}=\rho_{11}^{f_2}= \frac {1}{2}\sin^{2}\frac {\theta}{2}$ and 
$\rho_{0}^{f_2}= \rho_{00}^{f_2}+ \rho_{01}^{f_2}=
\rho_{1}^{f_2}=\rho_{10}^{f_2}+\rho_{11}^{f_2}=\frac {1}{2}$ are depicted with the solid 
lines. (a) and (b) $\theta=90^{o}$, (c) and (d) $\theta=53.24^{o}$.

Fig.4~ Coherence vs. $\phi$ when two paths are distinguishable. Data points 
$\bigcirc $ and $\bigtriangledown$ denote the coherence between the states $|00>$ and $|01>$, 
$*$ and $+$, between the states $|10>$ and $|11>$. Theoretical curves
expressed by $\pm \frac {1}{2}\sin\theta\sin\phi$ are depicted with the solid lines. (a) $\theta=90^{o}$, (b) $\theta=53.24^{o}$.

\end{document}